\documentclass[11pt,en,cite=authoryear]{elegantpaper}

\title{Construction of the Global $\chi^2$ Function for the Simultaneous Fitting of Correlated Energy-Dependent Cross Sections}
\author{Linquan Shao, Haoyu Yan, Yingjun Chen, Jiaxin Pi, Xingyu Zhou\footnote{Corresponding author: zhouxy@lnnu.edu.cn}}
\institute{School of Physics and Electronic Technology, Liaoning Normal University, Dalian 116029, China}

\usepackage{array}
\usepackage{bm}

\setcitestyle{numbers,super,square}

\begin{document}

\maketitle

\begin{abstract}
In this paper, the global $\chi^2$ function for the simultaneous fitting of correlated energy-dependent cross sections is constructed, where the correlations between the measured cross sections of different processes and/or at different center-of-mass energy points, as well as the contributions from the integrated luminosity measurement and the center-of-mass energy measurement, are taken into account.
\end{abstract}

\section{Introduction}

In high energy physics experiments, extracting the resonant parameters of hadrons, such as the $J/\psi$ meson, by fitting energy-dependent cross sections is a class of studies of fundamental importance\cite{BES,BaBar,CLEO,KEDR,KEDR2,KEDRpsip}. In such a study, to improve the precision, sometimes it is necessary to simultaneously fit multiple physical processes to extract the resonant parameters, such as mass and total width. In the simultaneous fits, all types of correlations among the involved cross sections have to be taken into account. Therefore, it becomes essential to construct a global $\chi^2$ function. In this paper, the global $\chi^2$ function for two physical processes are constructed with standard covariance matrix method.

\section{Construction of the global $\chi^2$ function}
\subsection{Known conditions}
Here, data taken at $n$ energy points are analyzed. Center-of-mass (CM) energies, integrated luminosities, cross sections of the two physics processes have been measured at all the $n$ energy points. For each of the last three quantities, not only $n$ values but also $n \times n$ covariance matrix has been obatined as follows,
\begin{align}
\begin{array}{ll}
L(i), \ \ \  \sigma_{\rm ch1}^{\rm exp}(i)), \ \ \  \sigma_{\rm ch2}^{\rm exp}(i) & i=1,2 \cdots n-1,n \\
V(L(i),L(j)) \equiv V_{L}(i,j) & i,j=1,2 \cdots n-1,n \\
V(\sigma_{\rm ch1}^{\rm exp}(i),\sigma_{\rm ch1}^{\rm exp}(j)) \equiv V_{\rm ch1}(i,j) & i,j=1,2 \cdots n-1,n \\
V(\sigma_{\rm ch2}^{\rm exp}(i),\sigma_{\rm ch2}^{\rm exp}(j)) \equiv V_{\rm ch2}(i,j) & i,j=1,2 \cdots n-1,n \\
\end{array}
\end{align}
As for the first quantities, namely CM energies, usually $n$ values and $n$ statistical errors are involved,
\begin{align}
\begin{array}{ll}
W(i) \pm \Delta W(i) & i=1,2 \cdots n-1,n \\
\end{array}
\end{align}
With them, we can construct a covariance matrix of the $n$ energies as follow,
\begin{equation}
V(W(i),W(j)) \equiv V_{W}(i,j) = \delta(i,j) (\Delta W(i))^2 \ \ \ i,j=1,2 \cdots n-1,n
\end{equation}
where
\begin{align}
\delta(i,j) = \left\{
\begin{array}{ll}
1 & i=j \\
0 & i \neq j \\
\end{array}
\right.
\end{align}

In addition, theoretical formulae of cross sections of two physics processes have also been obtained. Using these formulae, we can get theoretical values and their corresponding first order partial derivative values of cross sections of the two physics processes at all the $n$ energy points as follows,
\begin{equation}
\sigma_{\rm ch1}^{\rm the}(i), \ \ \  \frac{\partial \sigma_{\rm ch1}^{\rm the}}{\partial W}(i), \ \ \  \sigma_{\rm ch2}^{\rm the}(i), \ \ \  \frac{\partial \sigma_{\rm ch2}^{\rm the}}{\partial W}(i), \ \ \  i=1,2 \cdots n-1,n
\end{equation}

In the following, we will use all these experimental quantities and theoretical quantities to construct a $2n \times 2n$ global $\chi^2$ function which will be used in the simultaneous fitting of the two physics processes.

\subsection{Method}

According to the least square principle\cite{lsp}, the $\chi^2$ function can be constructed as follow,
\begin{equation}
\chi^2 = \bm{\Delta\sigma}^{\rm T} \cdot \bm{V}^{-1} \cdot \bm{\Delta\sigma}
\label{chisquare}
\end{equation}
where
\begin{equation}
\Delta\sigma(i) = \sigma^{\rm exp}(i) - \sigma^{\rm the}(i) \ \ \ \ \  i=1,2 \cdots 2n-1,2n
\label{Deltasigma1}
\end{equation}
To be specific,
\begin{equation}
\Delta\sigma(i) = 
\left\{
\begin{array}{ll}
\sigma_{\rm ch1}^{\rm exp}(i) - \sigma_{\rm ch1}^{\rm the}(i) & i=1,2 \cdots n-1,n \\
\sigma_{\rm ch2}^{\rm exp}(i-n) - \sigma_{\rm ch2}^{\rm the}(i-n) & i=n+1,n+2 \cdots 2n-1,2n \\
\end{array}
\right.
\label{Deltasigma2}
\end{equation}
If the CM energies are measured so precise that their uncertainties can be neglected, the element of the global covariance matrix, $V(i,j)$, can be evaluated simply as follow,
\begin{equation}
V(i,j) \equiv V(\sigma^{\rm exp}(i),\sigma^{\rm exp}(j)) \ \ \ \ \  i,j=1,2 \cdots 2n-1,2n
\end{equation}
Unfortunately, this is not the case in most analyses. Thus, we have to consider uncertainties of CM energy measurement. Once the uncertainties of energy measurement are considered, the theoretical cross sections $\sigma^{\rm the}(i)$ are no longer purely theoretical but become semi-experimental. In this case, we regard $\sigma^{\rm exp}(i) - \sigma^{\rm the}(i)$ as experimental quantities and 0 as their theoretical values, thus we can evaluate the element of the covariance matrix as follow,
\begin{equation}
V(i,j) \equiv V(\sigma^{\rm exp}(i) - \sigma^{\rm the}(i),\sigma^{\rm exp}(j) - \sigma^{\rm the}(j)) \ \ \ \ \  i,j=1,2 \cdots 2n-1,2n
\end{equation}
This is the fundamental formula for evaluating the element of the covariance matrix $V(i,j)$. In the following we will evaluate it according to the regions to which $i$ and $j$ belong.

\subsection{Derivation of the global covariance matrix elements}

For the case of $i=1,2 \cdots n-1,n, \  j=1,2 \cdots n-1,n$,
\begin{equation}
\sigma^{\rm exp}(i) =  \sigma_{\rm ch1}^{\rm exp}(i), \ \ \  \sigma^{\rm the}(i) = \sigma_{\rm ch1}^{\rm the}(i), \ \ \  \sigma^{\rm exp}(j) = \sigma_{\rm ch1}^{\rm exp}(j), \ \ \  \sigma^{\rm the}(j) = \sigma_{\rm ch1}^{\rm the}(j)
\end{equation}
thus
\begin{align}
\begin{array}{ll}
   & V(i,j) \\
 = & V(\sigma_{\rm ch1}^{\rm exp}(i) - \sigma_{\rm ch1}^{\rm the}(i),\sigma_{\rm ch1}^{\rm exp}(j) - \sigma_{\rm ch1}^{\rm the}(j)) \\
 = & V(\sigma_{\rm ch1}^{\rm exp}(i),\sigma_{\rm ch1}^{\rm exp}(j)) + V(\sigma_{\rm ch1}^{\rm the}(i),\sigma_{\rm ch1}^{\rm the}(j)) \\
 = & \displaystyle V_{\rm ch1}(i,j) + \sum \limits_{k=1}^{n} \sum \limits_{l=1}^{n} \frac{\partial \sigma_{\rm ch1}^{\rm the}(i)}{\partial W(k)} \frac{\partial \sigma_{\rm ch1}^{\rm the}(j)}{\partial W(l)} V(W(k),W(l)) \\
 = & \displaystyle V_{\rm ch1}(i,j) + \sum \limits_{k=1}^{n} \sum \limits_{l=1}^{n} \frac{\partial \sigma_{\rm ch1}^{\rm the}(i)}{\partial W(k)} \frac{\partial \sigma_{\rm ch1}^{\rm the}(j)}{\partial W(l)} \delta(k,l) (\Delta W(k))^2 \\
 = & \displaystyle V_{\rm ch1}(i,j) + \sum \limits_{k=1}^{n} \frac{\partial \sigma_{\rm ch1}^{\rm the}(i)}{\partial W(k)} \frac{\partial \sigma_{\rm ch1}^{\rm the}(j)}{\partial W(k)} (\Delta W(k))^2 \\
 = & \displaystyle V_{\rm ch1}(i,j) + \sum \limits_{k=1}^{n} \delta(k,i) \frac{\partial \sigma_{\rm ch1}^{\rm the}(i)}{\partial W(i)} \delta(k,j) \frac{\partial \sigma_{\rm ch1}^{\rm the}(j)}{\partial W(j)} (\Delta W(k))^2 \\
 = & \displaystyle V_{\rm ch1}(i,j) + \frac{\partial \sigma_{\rm ch1}^{\rm the}(i)}{\partial W(i)} \delta(i,j) \frac{\partial \sigma_{\rm ch1}^{\rm the}(j)}{\partial W(j)} (\Delta W(i))^2 \\
 = & \displaystyle V_{\rm ch1}(i,j) + \delta(i,j) \left(\frac{\partial \sigma_{\rm ch1}^{\rm the}(i)}{\partial W(i)} \Delta W(i)\right)^2 \\
 = & \displaystyle V_{\rm ch1}(i,j) + \delta(i,j) \left(\frac{\partial \sigma_{\rm ch1}^{\rm the}}{\partial W}(i) \Delta W(i)\right)^2 \\
\end{array}
\end{align}
In the result, the first item is exactly the covariance matrix element of the first-channel cross section measurement, and the second item contributes from the uncertainty of the first-channel theoretical crosss section due to the CM energy measurement.

For the case of $i=1,2 \cdots n-1,n, \  j=n+1,n+2 \cdots 2n-1,2n$,
\begin{equation}
\sigma^{\rm exp}(i)=\sigma_{\rm ch1}^{\rm exp}(i), \ \ \  \sigma^{\rm the}(i)=\sigma_{\rm ch1}^{\rm the}(i), \ \ \  \sigma^{\rm exp}(j)=\sigma_{\rm ch2}^{\rm exp}(j-n), \ \ \  \sigma^{\rm the}(j)=\sigma_{\rm ch2}^{\rm the}(j-n)
\end{equation}
thus
\begin{align}
\begin{array}{ll}
   & V(i,j) \\
 = & V(\sigma_{\rm ch1}^{\rm exp}(i) - \sigma_{\rm ch1}^{\rm the}(i),\sigma_{\rm ch2}^{\rm exp}(j-n) - \sigma_{\rm ch2}^{\rm the}(j-n)) \\
 = & V(\sigma_{\rm ch1}^{\rm exp}(i),\sigma_{\rm ch2}^{\rm exp}(j-n)) + V(\sigma_{\rm ch1}^{\rm the}(i),\sigma_{\rm ch2}^{\rm the}(j-n)) \\
 = & \displaystyle V(\frac{N_{\rm ch1}(i)}{L_(i) \cdot \epsilon_{\rm ch1}(i)},\frac{N_{\rm ch2}(j-n)}{L(j-n) \cdot \epsilon_{\rm ch2}(j-n)}) + \sum \limits_{k=1}^{n} \sum \limits_{l=1}^{n} \frac{\partial \sigma_{\rm ch1}^{\rm the}(i)}{\partial W(k)} \frac{\partial \sigma_{\rm ch2}^{\rm the}(j-n)}{\partial W(l)} V(W(k),W(l)) \\
 = & \displaystyle \sum \limits_{k=1}^{n} \sum \limits_{l=1}^{n} \frac{\partial \sigma_{\rm ch1}^{\rm exp}(i)}{\partial L(k)} \frac{\partial \sigma_{\rm ch2}^{\rm exp}(j-n)}{\partial L(l)} V(L(k),L(l)) + \sum \limits_{k=1}^{n} \sum \limits_{l=1}^{n} \frac{\partial \sigma_{\rm ch1}^{\rm the}(i)}{\partial W(k)} \frac{\partial \sigma_{\rm ch2}^{\rm the}(j-n)}{\partial W(l)} \delta(k,l) (\Delta W(k))^2 \\
 = & \displaystyle \sum \limits_{k=1}^{n} \sum \limits_{l=1}^{n} \delta(k,i) (-\frac{\sigma_{\rm ch1}^{\rm exp}(i)}{L(k)}) \delta(j,j-n) (-\frac{\sigma_{\rm ch2}^{\rm exp}(j-n)}{L(l)}) V_{L}(k,l) + \sum \limits_{k=1}^{n} \frac{\partial \sigma_{\rm ch1}^{\rm the}(i)}{\partial W(k)} \frac{\partial \sigma_{\rm ch2}^{\rm the}(j-n)}{\partial W(k)} (\Delta W(k))^2 \\
 = & \displaystyle \frac{\sigma_{\rm ch1}^{\rm exp}(i)}{L(i)} \frac{\sigma_{\rm ch2}^{\rm exp}(j-n)}{L(j-n)} V_{L}(i,j-n) + \sum \limits_{k=1}^{n} \delta(k,i) \frac{\partial \sigma_{\rm ch1}^{\rm the}(i)}{\partial W(i)} \delta(k,j-n) \frac{\partial \sigma_{\rm ch2}^{\rm the}(j-n)}{\partial W(j-n)} (\Delta W(k))^2 \\
 = & \displaystyle \frac{\sigma_{\rm ch1}^{\rm exp}(i) \, \sigma_{\rm ch2}^{\rm exp}(j-n)}{L(i) \, L(j-n)} V_{L}(i,j-n) + \frac{\partial \sigma_{\rm ch1}^{\rm the}(i)}{\partial W(i)} \delta(i,j-n) \frac{\partial \sigma_{\rm ch2}^{\rm the}(j-n)}{\partial W(j)} (\Delta W(i))^2 \\
 = & \displaystyle \frac{\sigma_{\rm ch1}^{\rm exp}(i) \, \sigma_{\rm ch2}^{\rm exp}(j-n)}{L(i) \, L(j-n)} V_{L}(i,j-n) + \delta(i,j-n) \frac{\partial \sigma_{\rm ch1}^{\rm the}(i)}{\partial W(i)} \frac{\partial \sigma_{\rm ch2}^{\rm the}(i)}{\partial W(i)} (\Delta W(i))^2 \\
 = & \displaystyle \frac{\sigma_{\rm ch1}^{\rm exp}(i) \, \sigma_{\rm ch2}^{\rm exp}(j-n)}{L(i) \, L(j-n)} V_{L}(i,j-n) + \delta(i,j-n) \frac{\partial \sigma_{\rm ch1}^{\rm the}}{\partial W}(i) \frac{\partial \sigma_{\rm ch2}^{\rm the}}{\partial W}(i) (\Delta W(i))^2 \\
\end{array}
\end{align}
In the result, the first item contributes from the covariance matrix element of the integrated luminosity measurement, and the second item contributes from the uncertainty of the first-channel and second-channel theoretical crosss sections due to the CM energy measurement.

For the case of $i=n+1,n+2 \cdots 2n-1,2n, \  j=1,2 \cdots n-1,n$,
\begin{equation}
\sigma^{\rm exp}(i)=\sigma_{\rm ch2}^{\rm exp}(i-n), \ \ \  \sigma^{\rm the}(i)=\sigma_{\rm ch2}^{\rm the}(i-n), \ \ \  \sigma^{\rm exp}(j)=\sigma_{\rm ch1}^{\rm exp}(j), \ \ \  \sigma^{\rm the}(j)=\sigma_{\rm ch1}^{\rm the}(j)
\end{equation}
thus
\begin{align*}
\begin{array}{ll}
   & V(i,j) \\
 = & V(\sigma_{\rm ch2}^{\rm exp}(i-n) - \sigma_{\rm ch2}^{\rm the}(i-n),\sigma_{\rm ch1}^{\rm exp}(j) - \sigma_{\rm ch1}^{\rm the}(j)) \\
 = & V(\sigma_{\rm ch2}^{\rm exp}(i-n),\sigma_{\rm ch1}^{\rm exp}(j)) + V(\sigma_{\rm ch2}^{\rm the}(i-n),\sigma_{\rm ch1}^{\rm the}(j)) \\
 = & \displaystyle V(\frac{N_{\rm ch2}(i-n)}{L_(i-n) \cdot \epsilon_{\rm ch2}(i-n)},\frac{N_{\rm ch1}(j)}{L(j) \cdot \epsilon_{\rm ch1}(j)}) + \sum \limits_{k=1}^{n} \sum \limits_{l=1}^{n} \frac{\partial \sigma_{\rm ch2}^{\rm the}(i-n)}{\partial W(k)} \frac{\partial \sigma_{\rm ch1}^{\rm the}(j)}{\partial W(l)} V(W(k),W(l)) \\
 = & \displaystyle \sum \limits_{k=1}^{n} \sum \limits_{l=1}^{n} \frac{\partial \sigma_{\rm ch2}^{\rm exp}(i-n)}{\partial L(k)} \frac{\partial \sigma_{\rm ch1}^{\rm exp}(j)}{\partial L(l)} V(L(k),L(l)) + \sum \limits_{k=1}^{n} \sum \limits_{l=1}^{n} \frac{\partial \sigma_{\rm ch2}^{\rm the}(i-n)}{\partial W(k)} \frac{\partial \sigma_{\rm ch1}^{\rm the}(j)}{\partial W(l)} \delta(k,l) (\Delta W(k))^2 \\
 = & \displaystyle \sum \limits_{k=1}^{n} \sum \limits_{l=1}^{n} \delta(k,i-n) (-\frac{\sigma_{\rm ch2}^{\rm exp}(i-n)}{L(k)}) \delta(j,n) (-\frac{\sigma_{\rm ch1}^{\rm exp}(j)}{L(l)}) V_{L}(k,l) + \sum \limits_{k=1}^{n} \frac{\partial \sigma_{\rm ch2}^{\rm the}(i-n)}{\partial W(k)} \frac{\partial \sigma_{\rm ch1}^{\rm the}(j)}{\partial W(k)} (\Delta W(k))^2 \\
 = & \displaystyle \frac{\sigma_{\rm ch2}^{\rm exp}(i-n)}{L(i-n)} \frac{\sigma_{\rm ch1}^{\rm exp}(j)}{L(j)} V_{L}(i-n,j) + \sum \limits_{k=1}^{n} \delta(k,i-n) \frac{\partial \sigma_{\rm ch2}^{\rm the}(i-n)}{\partial W(i-n)} \delta(k,j) \frac{\partial \sigma_{\rm ch1}^{\rm the}(j)}{\partial W(j)} (\Delta W(k))^2 \\
\end{array}
\end{align*}
\begin{align}
\begin{array}{ll}
 = & \displaystyle \frac{\sigma_{\rm ch2}^{\rm exp}(i-n) \, \sigma_{\rm ch1}^{\rm exp}(j)}{L(i-n) \, L(j)} V_{L}(i-n,j) + \frac{\partial \sigma_{\rm ch2}^{\rm the}(i-n)}{\partial W(i-n)} \delta(i-n,j) \frac{\partial \sigma_{\rm ch1}^{\rm the}(j)}{\partial W(j)} (\Delta W(i-n))^2 \\
 = & \displaystyle \frac{\sigma_{\rm ch2}^{\rm exp}(i-n) \, \sigma_{\rm ch1}^{\rm exp}(j)}{L(i-n) \, L(j)} V_{L}(i-n,j) + \delta(i-n,j) \frac{\partial \sigma_{\rm ch2}^{\rm the}(j)}{\partial W(j)} \frac{\partial \sigma_{\rm ch1}^{\rm the}(j)}{\partial W(j)} (\Delta W(j))^2 \\
 = & \displaystyle \frac{\sigma_{\rm ch2}^{\rm exp}(i-n) \, \sigma_{\rm ch1}^{\rm exp}(j)}{L(i-n) \, L(j)} V_{L}(i-n,j) + \delta(i-n,j) \frac{\partial \sigma_{\rm ch2}^{\rm the}}{\partial W}(j) \frac{\partial \sigma_{\rm ch1}^{\rm the}}{\partial W}(j) (\Delta W(j))^2 \\
\end{array}
\end{align}
In the result, the first item contributes from the covariance matrix element of the integrated luminosity measurement, and the second item contributes from the uncertainty of the first-channel and second-channel theoretical crosss sections due to the CM energy measurement.

For the case of $i=n+1,n+2 \cdots 2n-1,2n, \  j=n+1,n+2 \cdots 2n-1,2n$,
\begin{equation}
\sigma^{\rm exp}(i)=\sigma_{\rm ch2}^{\rm exp}(i-n), \ \ \  \sigma^{\rm the}(i)=\sigma_{\rm ch2}^{\rm the}(i-n), \ \ \  \sigma^{\rm exp}(j)=\sigma_{\rm ch2}^{\rm exp}(j-n), \ \ \  \sigma^{\rm the}(j)=\sigma_{\rm ch2}^{\rm the}(j-n)
\end{equation}
thus
\begin{align}
\begin{array}{ll}
   & V(i,j) \\
 = & V(\sigma_{\rm ch2}^{\rm exp}(i-n) - \sigma_{\rm ch2}^{\rm the}(i-n),\sigma_{\rm ch2}^{\rm exp}(j-n) - \sigma_{\rm ch2}^{\rm the}(j-n)) \\
 = & V(\sigma_{\rm ch2}^{\rm exp}(i-n),\sigma_{\rm ch2}^{\rm exp}(j-n)) + V(\sigma_{\rm ch2}^{\rm the}(i-n),\sigma_{\rm ch2}^{\rm the}(j-n)) \\
 = & \displaystyle V_{\rm ch2}(i-n,j-n) + \sum \limits_{k=1}^{n} \sum \limits_{l=1}^{n} \frac{\partial \sigma_{\rm ch2}^{\rm the}(i-n)}{\partial W(k)} \frac{\partial \sigma_{\rm ch2}^{\rm the}(j-n)}{\partial W(l)} V(W(k),W(l)) \\
 = & \displaystyle V_{\rm ch2}(i-n,j-n) + \sum \limits_{k=1}^{n} \sum \limits_{l=1}^{n} \frac{\partial \sigma_{\rm ch2}^{\rm the}(i-n)}{\partial W(k)} \frac{\partial \sigma_{\rm ch2}^{\rm the}(j-n)}{\partial W(l)} \delta(k,l) (\Delta W(k))^2 \\
 = & \displaystyle V_{\rm ch2}(i-n,j-n) + \sum \limits_{k=1}^{n} \frac{\partial \sigma_{\rm ch2}^{\rm the}(i-n)}{\partial W(k)} \frac{\partial \sigma_{\rm ch2}^{\rm the}(j-n)}{\partial W(k)} (\Delta W(k))^2 \\
 = & \displaystyle V_{\rm ch2}(i-n,j-n) + \sum \limits_{k=1}^{n} \delta(k,i-n) \frac{\partial \sigma_{\rm ch2}^{\rm the}(i-n)}{\partial W(i-n)} \delta(k,j-n) \frac{\partial \sigma_{\rm ch2}^{\rm the}(j-n)}{\partial W(j-n)} (\Delta W(k))^2 \\
 = & \displaystyle V_{\rm ch2}(i-n,j-n) + \frac{\partial \sigma_{\rm ch2}^{\rm the}(i-n)}{\partial W(i-n)} \delta(i,j) \frac{\partial \sigma_{\rm ch2}^{\rm the}(j-n)}{\partial W(j-n)} (\Delta W(i-n))^2 \\
 = & \displaystyle V_{\rm ch2}(i-n,j-n) + \delta(i,j) \left(\frac{\partial \sigma_{\rm ch2}^{\rm the}(i-n)}{\partial W(i-n)} \Delta W(i-n)\right)^2 \\
 = & \displaystyle V_{\rm ch2}(i-n,j-n) + \delta(i,j) \left(\frac{\partial \sigma_{\rm ch2}^{\rm the}}{\partial W}(i-n) \Delta W(i-n)\right)^2 \\
\end{array}
\end{align}
In the result, the first item is exactly the covariance matrix element of the second-channel cross section measurement, and the second item contributes from the uncertainty of the second-channel theoretical crosss section due to the CM energy measurement.

\subsection{Result}

Sum up the results in the four regions, we obtain
\begin{equation}
V(i,j) =
\left\{
\begin{array}{ll}
\displaystyle V_{\rm ch1}(i,j) + \delta(i,j) \, \left(\frac{\partial \sigma_{\rm ch1}^{\rm the}}{\partial W}(i) \, \Delta W(i)\right)^2 & \displaystyle {\Large{\textcircled{1}}} \\
\displaystyle \frac{\sigma_{\rm ch1}^{\rm exp}(i) \, \sigma_{\rm ch2}^{\rm exp}(j-n)}{L(i) \, L(j-n)} \, V_{L}(i,j-n) + \delta(i,j-n) \, \frac{\partial \sigma_{\rm ch1}^{\rm the}}{\partial W}(i) \, \frac{\partial \sigma_{\rm ch2}^{\rm the}}{\partial W}(i) \, (\Delta W(i))^2 & \displaystyle {\Large{\textcircled{2}}} \\
\displaystyle \frac{\sigma_{\rm ch2}^{\rm exp}(i-n) \, \sigma_{\rm ch1}^{\rm exp}(j)}{L(i-n) \, L(j)} V_{L}(i-n,j) + \delta(i-n,j) \, \frac{\partial \sigma_{\rm ch2}^{\rm the}}{\partial W}(j) \, \frac{\partial \sigma_{\rm ch1}^{\rm the}}{\partial W}(j) \, (\Delta W(j))^2 & \displaystyle {\Large{\textcircled{3}}} \\
\displaystyle V_{\rm ch2}(i-n,j-n) + \delta(i,j) \, \left(\frac{\partial \sigma_{\rm ch2}^{\rm the}}{\partial W}(i-n) \, \Delta W(i-n)\right)^2 & \displaystyle {\Large{\textcircled{4}}} \\
\end{array}
\right.
\label{Vij}
\end{equation}
with
\begin{equation}
\left\{
\begin{array}{l}
\displaystyle {\Large{\textcircled{1}}} \text{ denotes } i=1,2 \cdots n-1,n, \  j=1,2 \cdots n-1,n \\
\displaystyle {\Large{\textcircled{2}}} \text{ denotes } i=1,2 \cdots n-1,n, \  j=n+1,n+2 \cdots 2n-1,2n \\
\displaystyle {\Large{\textcircled{3}}} \text{ denotes } i=n+1,n+2 \cdots 2n-1,2n, \  j=1,2 \cdots n-1,n \\
\displaystyle {\Large{\textcircled{4}}} \text{ denotes } i=n+1,n+2 \cdots 2n-1,2n, \  j=n+1,n+2 \cdots 2n-1,2n \\
\end{array}
\right.
\label{Rangesij}
\end{equation}

\section{Summary}

To take into account the correlations between the measured cross sections of different processes and/or at different CM energy points, as well as the contributions from the integrated luminosity measurement and the CM energy measurement, we construct a global $\chi^2$ function for the simultaneous fitting of correlated energy-dependent cross sections, according to the standard covariance matrix method.
The result is shown in the formulae (\ref{chisquare}), (\ref{Deltasigma2}), (\ref{Vij}), and (\ref{Rangesij}).  This result has already been used in the measurement of the total and leptonic decay widths of the $J/\psi$ resonanceat at BESIII\cite{BESIII}. In addition, though the result only applies to the case of two channels, however, it can be generized to the cases of three or more channels easily with the same method.

\end{document}